\documentclass[twocolumn,showpacs,preprintnumbers,amsmath]{revtex4}

\usepackage{graphicx}
\usepackage{dcolumn}
\usepackage{bm}
\usepackage{amsmath}

\newcommand{\dd}{\mathrm{d}}
\newcommand{\dpt}{\partial}

\begin{document}

\preprint{APS/123-QED}

\title{Dewetting of thin-film polymers}

\author{F. Saulnier} \email{florent.saulnier@college-de-france.fr}
\author{E. Rapha\"{e}l} \email{elie.raphael@college-de-france.fr}
\author{P.-G. de Gennes} \email{pgg@espci.fr}

\affiliation{%
Laboratoire de Physique de la Mati\`ere Condens\'ee, CNRS UMR 7125,
Coll\`ege de France\\
11, place Marcelin Berthelot, 75231 Paris Cedex 05, France.}

\date{21th of june, 2002}

\begin{abstract}
In this paper we present a theoretical model for the dewetting of
ultra-thin polymer films. Assuming that the shear-thinning
properties of these films can be described by a Cross-type
constitutive equation, we analyze the front morphology of the
dewetting film, and characterize the time evolution of the dry
region radius, and of the rim height. Different regimes of growth
are expected, depending on the initial film thickness, and on the
power-law index involved in the constitutive equation. In the
thin-films regime, the dry radius and the rim height obey
power-law time dependences. We then compare our predictions with
the experimental results obtained by Debr\'egeas {\it et al.}
[Phys. Rev. Lett. {\bf 75}, 3886 (1995)] and by Reiter [Phys. Rev.
Lett. {\bf 87}, 186101 (2001)].

\end{abstract}


\maketitle

\section{Introduction}

Thin liquid films have considerable scientifical and technological
importance, and have numerous applications. In engineering, for
instance, they serve to protect surfaces, and applications arise
in paints, adhesives and membranes \cite{oron}. Thin liquids films
display a variety of interesting dynamics phenomena and have
therefore been the focus of many experimental and theoretical
studies \cite{revues}. In particular, thin polymer films have
recently attracted a lot of interest since understanding their
properties such as viscosity \cite{masson}, chain mobility
\cite{forrest} and stability \cite{reiterpgg} is essential for
optimization.

A long time ago, Taylor \cite{taylor} and Culick \cite{culick}
analyzed the growth of a circular hole in a thin liquid sheet
\cite{dupre}. By balancing surface tension forces against inertia
\cite{keller1}, they found that the rim of liquid at the edge of
the films retracts at a constant velocity, a prediction first
checked experimentally by Mc Entee and Mysels
\cite{mysels,keller}. Debr\'egeas and collaborators
\cite{debregeas} have recently studied the bursting of thin
suspended films of very viscous liquids. These experiments
revealed unexpected features: (a) First, the retraction velocity
grows exponentially with time (with a characteristic time scale
$\tau_{i} = h_{i} \eta /|S|$, where $h_i$, $\eta$ and $S$ are
respectively the initial film thickness, the viscosity and the
spreading coefficient \cite{rem1}), (b) second, the liquid is {\it
not} collected into a rim and the film remains flat through the
retraction. According to these authors, the uniform thickening of
the retracting film was a consequence of its viscoelasticity,
which permits an elastic propagation into the film of the surface
tension forces acting on the edge. Brenner and Gueyffier
\cite{brenner} showed, however, that the absence of rim can also
result from a purely viscous effect \cite{koplik,liu}.

Very recently \cite{reiter}, Reiter studied the dewetting of
ultrathin ({\it i.e.} thinner than the coil size), almost glassy
polystyrene (PS) films deposited onto silicon wafers coated with a
polydimethylsiloxane (PDMS) monolayer. He found that a highly
asymmetric rim, with an extremely steep side towards the interior
of the hole and a much slower decay on the rear side, builds up
progressively \cite{herminghaus}, with the maximum height
increasing linearly with the diameter of the hole. We recently
\cite{saulnier} proposed a theoretical model to explain such
deviations from the behavior of pure liquids, based on the
shear-thinning properties of the polymer film. Assuming that the
stresses inside the film saturate logarithmically with the strain
rates, we showed that different regimes of growth are expected,
depending on the initial film thickness and the experimental time
range. Other theoretical approaches for polymer film dewetting
have recently been proposed by Herminghaus {\it et al.}
\cite{condmat} and Shenoy {\it et al} \cite{shenoy}.

In this paper we aim at precising our methods of resolution, and
characterizing the dewetting process for a Cross-type constitutive
equation (interpolation between a viscous behavior at low
strain-rates and a power-law dependence for the viscosity versus
the shear strain rate), following recent experimental results by
Dalnoki-Veress {\it et al.} for PS films \cite{dalnoki}. We study
the profile of the dewetting film, and characterize the time
evolution of the dry radius, and of the rim height (as summarized
in appendix \ref{bilan}). Finally, we discuss the adequation
between our theoretical predictions and experimental results by
Debr\'egeas and Reiter.

\section{The model}
\label{the model}
\begin{figure}
\resizebox{0.40 \textwidth}{!}{%
  \includegraphics*[3.4cm,11cm][18cm,18.2cm]{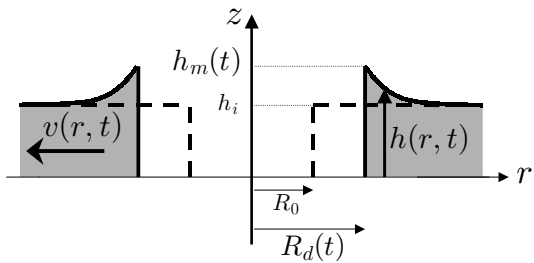} }
\caption{Film geometry: $h(r,t)$ is the profile of the film,
$h_{m}(t)$ is the height of the rim, and $R_{d}(t)$ is the radius
of the dry zone. The initial ($t = 0$) step-like profile is
represented by the dashed line. $v(r,t)$ is the radial,
axisymmetric flow field.} \label{dessin modele}
\end{figure}

Figure \ref{dessin modele} shows the film
geometry. $h(r,t)$ is the profile of the film, $h_{m}(t)$ is the
height of the rim, and $R_{d}(t)$ is the radius of the dry zone.
$v(r,t)$ is the radial, axisymmetric flow field (cf. appendix \ref{vz}). On a
non-wettable, smooth and passive solid substrate like the
PDMS-coated silicon wafer used by Reiter, this plug-flow
description is valid as long as $h_{i} \ll b$ where $b$ is the
hydrodynamic extrapolation length (cf. \cite{rem2}). The range of
thicknesses studied by Reiter allows this simplifying assumption.

\subsection{Constitutive law of the material}

In order to characterize the rheologic properties of the film, we
introduce the stress tensor $\sigma_{ij}=-p \delta_{ij}+\sigma_{\:
\: \: ij}^{m}$, where $p$ is the pressure and $\sigma_{\: \: \:
ij}^{m}$ represents the effects of internal friction. We relate
$\sigma_{\: \: \: ij}^{m}$ to the strain rate tensor
$\dot{\gamma}_{ij}$ by a constitutive law of the form:

\begin{equation}
\sigma^{m} (\dot{\gamma}) = \sigma_{0} \: \Phi (\dot{\gamma} \tau),
\label{rheolaw}
\end{equation}

where $\sigma_{0}$ and $\tau$ are material constants, and $\Phi$
is a generic function.

For a purely viscous liquid, the function
$\Phi$ is linear:

\begin{equation}
\Phi(\dot{\gamma} \tau) = \dot{\gamma} \tau
\label{viscous}
\end{equation}

In this case, the fluid viscosity $\eta = \sigma^m / \dot{\gamma}$
is constant and equals the zero-shear viscosity
$\eta_{0}=\sigma_{0} \tau$.

For polymers, just above $T_{g}$, it is expected within the
framework of the free-volume model that $\sigma^{m}$ is expected
to vary logarithmically with ${\dot{\gamma}}$ as \cite{dyre}:

\begin{equation}
\Phi(\dot{\gamma} \tau) = \ln(1+\dot{\gamma} \tau)
\label{shear-thinning}
\end{equation}

The features of the dewetting regimes obtained with this
rheological law were presented elsewhere \cite{saulnier}, and will
only be briefly mentioned in the present paper in Appendix
\ref{log}. As mentioned in \cite{saulnier}, this logarithmic law does not
permit a simple analytical description of {\it all} the regimes of
time and thicknesses.

An alternative way of describing the polymer rheology is to use
the well-known Cross model \cite{cross}:
\begin{equation}
\Phi(\dot{\gamma} \tau) =\frac{\dot{\gamma} \tau}{1+k (\dot{\gamma}
\tau)^n},
\label{cross model}
\end{equation}
$k$ and $n$ being dimensionless constant parameters. A {\it
shear-thinning behavior} is taken into account by taking the
power-law index $n$ strictly between 0 and 1. At low strain-rates
($\dot{\gamma} \tau <k^{-1/n}$), this law displays a viscous-type
behavior ($\sigma^{m} \approx \eta_{0} \dot{\gamma}$, with a
zero-shear viscosity $\eta_{0}=\sigma_{0} \tau$), while for large
values of $\dot{\gamma} \tau$, $\sigma^{m}$ obeys the well-known
power-law model. This popular expression, which will be used all
along with this article, is applicable to a number of polymer
materials and complex fluids \cite{barnes}.

\subsection{Equations}
Assuming the fluid to be incompressible, mass conservation leads
to:

\begin{equation}
\frac{1}{h(r,t)} (\frac{\partial h(r,t)}{\partial
t}+\frac{r \beta(r,t)}{\tau} \frac{\partial h(r,t)}{\partial
r})=\frac{\alpha(r,t)-\beta(r,t)}{\tau} \label{eqn hrt}
\end{equation}

Equation \ref{eqn hrt} involves two unknown functions
$\alpha(r,t)$ and $\beta(r,t)$, that are positive, dimensionless
forms of the strain-rate components $\dot{\gamma}_{rr}$ and
$\dot{\gamma}_{\phi \phi}$:

\begin{equation}
\left\{ \begin{array}{ll} \alpha = -\tau\dot{\gamma}_{rr}=-\tau
\frac{\partial
v}{\partial r} \\
\beta = \tau \dot{\gamma}_{\phi \phi}=\tau \frac{v}{r}
                    \end{array}
            \right.
\label{definition}
\end{equation}

We thus need two additional equations to determine $h(r,t)$.
First, note that the following partial differential equation can
be directly derived from Eq.\ref{definition}:

\begin{equation}
\frac{\partial \beta}{\partial r}= - \frac{\beta +\alpha}{r}
\label{eqn suppl}
\end{equation}

Neglecting the inertial term, conservation of momentum
(projected on the radial direction \cite{christensen}) leads to:

\begin{equation}
\frac{\partial \sigma_{rr}}{\partial r}
+\frac{\sigma_{rr}-\sigma_{\phi \phi}}{r}=0
\label{motion}
\end{equation}

\begin{figure}
\resizebox{0.45 \textwidth}{!}{%
  \includegraphics*[1cm,9.5cm][19cm,22cm]{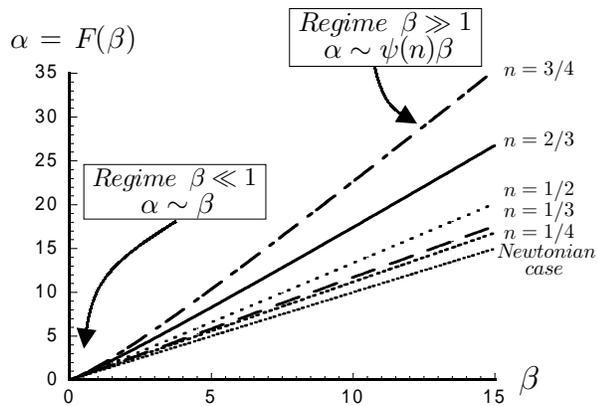} }
\caption{The radial strain-rate, $\alpha$, and the orthoradial
strain-rate, $\beta$, are related by $\alpha=F(\beta)$ (see
Eq.\ref{f(beta)}). In the purely Newtonian case
($\Phi(\dot{\gamma}\tau) =\dot{\gamma}\tau$), the function $F$ is
linear: $\alpha=\beta$. The other curves represent $F$ for the
Cross model (Eq.\ref{cross model}), taking different values of $n$
for a given value of $k=10$: $n=1/4; 1/3; 1/2; 2/3$ (an important
case, as precised in part \ref{gunter exp}); and $n=3/4$. The two
limit regimes $\beta \ll 1$ and $\beta \gg 1$, and the
corresponding approached expressions for $F$ are also
represented.} \label{function f(beta)}
\end{figure}

It can be shown that Eqs. \ref{eqn suppl} and \ref{motion}, along
with the free-surface boundary condition ({\it i.e.}
$\sigma_{zz}=0$ at the contact with ambient atmosphere
\cite{rem4}), allow one to express the strain rate $\alpha$ as a
function of $\beta$ only : $\alpha=F(\beta)$. Substitution of the
shear-thinning constitutive law \ref{rheolaw} in Eq.\ref{motion}
then leads to the following differential equation for $F$:

\begin{equation}
 \frac{\mbox{d}[\Phi(F(\beta))-\Phi(F(\beta)-\beta)]}{\mbox{d} \beta} =
\frac{\Phi(F(\beta))-\Phi(\beta)}{F(\beta) + \beta},
\label{f(beta)}
\end{equation}

which has to be solved along with the condition $F(0)=0$ (far away
from the perturbed central region, the strain rates must decrease
to zero). In Fig.\ref{function f(beta)}, we present the form of
functions $\alpha=F(\beta)$ for the Cross model, taking for each
curve $k=10$ and different values of power-law index $n$ between 0
and 1. This function can be approached by simpler expressions in
the two opposite regimes $\beta \ll 1$ and $\beta \gg 1$, as
explained below (cf.\ref{thin initial}).

In order to solve Eq.\ref{eqn hrt}, we should supply it with
initial and boundary conditions. Our initial profile is assumed to
be uniform, with a thickness $h_i$, except in a bored region
ranging from $r=0$ to $r=R_{0}$ (as shown in fig.\ref{dessin
modele}). $R_{0}$ is our characteristic radial length used
thereafter to make $r$ dimensionless:

\begin{equation}
h(r,t=0)= \left\{ \begin{array}{ll} h_{i} & \mbox{if $r \geq R_{0}=1$} \\
                                    0 & \mbox{otherwise}
                    \end{array}
            \right.
\label{boundary}
\end{equation}

We do not consider here the origin of the initial dewetting
process: experimentally, it is found that a thick PS film on a
silicon substrate is metastable and dewets via nucleation and
growth of dry patches \cite{redon}, while thinner films
($h_{i}<$100 nm) are unstable and dewet by spinodal decomposition
\cite{gunterprec,safran}. In his latest experiments on ultra-thin films
\cite{reiter}, Reiter characterized the early stage of the
dewetting process by the formation and coalescence of little
holes, with the displaced material uniformly distributed between
the holes, without visible rims. Our initial time $t=0$ might
correspond to the end of this preliminary process.

Equation \ref{eqn hrt} applies outside the dewetted region
($R_{d}(t) \leq r< \infty$). At the edge of the rim
($r=R_{d}(t)$), the rim height, $h_{m}(t)$, can be determined by
taking into account capillary forces. The applied force on the
rim, pushing the film away the dry area, must be balanced by the
internal radial stress: $|S|=|\sigma_{rr}| h_{m}(t)$. Assuming the
lateral extension of the film to be large enough, the film
thickness must reach $h_{i}$ far from the dry region: $\lim_{r
\rightarrow \infty}h(r,t)=h_{i} \; \; (\forall t)$. The complete
resolution of our set of equations \ref{eqn hrt}-\ref{eqn
suppl}-\ref{f(beta)} can be achieved using a method of
characteristics (cf. appendix \ref{appa}).

Thereafter, all thicknesses will be made dimensionless by
normalizing with a characteristic length $h^{*} \equiv
|S|/\sigma_{0}$. For Reiter's experimental conditions, at $T
\approx 105^{\circ}$C, we estimate $h^{*}$ to be of the order of
500 \AA. This parameter $h^{*}$ is of crucial importance in our
model, because it discriminates two regimes of growth whose
features are quite different. The films whose initial thickness
$h_i$ is larger than $h^{*}$ will be considered as "thick" films, while
the others ($h_i \ll h^{*}$) will be the "thin" ones.

\section{Results}

Note first that for a purely viscous liquid ($\Phi (\dot{\gamma}
\tau)=\dot{\gamma} \tau$), our model leads to a constant and
uniform thickness for the film, with an exponential growth of the
dry radius:

\begin{equation}
\left\{ \begin{array}{ll} R_{d}(t) = R_{0}
e^{\frac{|S|t}{\sigma_{0} \tau h_{i}} } = R_{0}
e^{\frac{t}{\tau_{i}}} \\
                          h_m(t) \approx h_i
                    \end{array}
            \right.
\label{reg visqueux}
\end{equation}

This is in complete agreement with the experimental results of
Debr\'egeas {\it et al.} for the dewetting of suspended polymer
films \cite{debregeas} and supported films
\cite{debregeasmacromol}.

How are affected the growth laws of the dry radius $R_{d}(t)$ and
the rim height $h_{m}(t)$ by the shear-thinning properties of the
material ?

In order to discuss the time dependence of the rim height and the
dry radius, we need to compare $h_{i}$ with the characteristic
thickness $h^{*}$.

\subsection{Thin films regime ($h_{i} \ll h^{*}$)}
\subsubsection{Initial stages of growth}
\label{thin initial}

\begin{figure}
\resizebox{0.47 \textwidth}{!}{%
\includegraphics*[3.5cm,10cm][19.5cm,22cm]{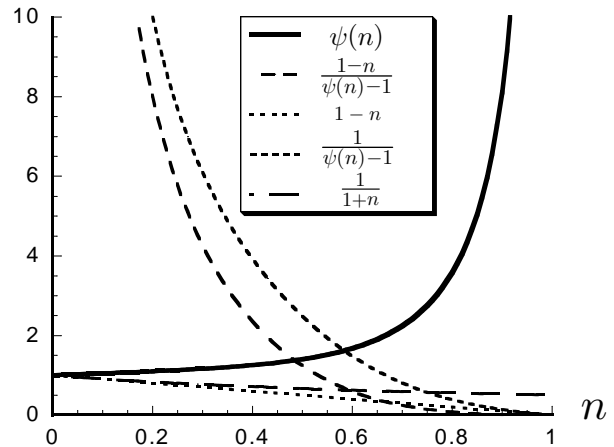} }
\caption{Variation with the power-law index $n$ of the parameter
$\psi(n)$. Some other quantities of interest are also represented
(see text).} \label{coeffs}
\end{figure}

The range $h_{i} \ll h^{*}$ allows different regimes of growth for
the initial stage of hole formation, depending on the specific
characteristics of the constitutive law.

Simple analytical laws can be obtained with the Cross law
($\Phi(\dot{\gamma} \tau) =(\dot{\gamma} \tau)/(1+k (\dot{\gamma}
\tau)^n)$). It can be shown that the solution $\alpha[\beta]$ of
Eq.\ref{f(beta)} is characterized by $\alpha \sim \beta$ for
low strain-rates, and $\alpha \sim \psi(n) \beta$ for $\beta \gg
1$. It is worth noting that the coefficient $\psi(n)$ does not
depend on $k$, but depends on $n$ as a solution of:
\begin{equation}
(\psi^{1-n}+(\psi-1)^{1-n})(\psi+1)(1-n)=\psi^{1-n}+1
\label{def of psi}
\end{equation}
As shown in Fig.\ref{coeffs}, this coefficient ranges
from $\psi = 1$ (for $n \rightarrow 0$) to $\psi \rightarrow
+\infty$ (for $n \rightarrow 1$), and is involved in the laws of
growth relative to the rim height and dry radius, as shown below.

\begin{figure}
\resizebox{0.45 \textwidth}{!}{%
\includegraphics*[1.8cm,8.2cm][19cm,21cm]{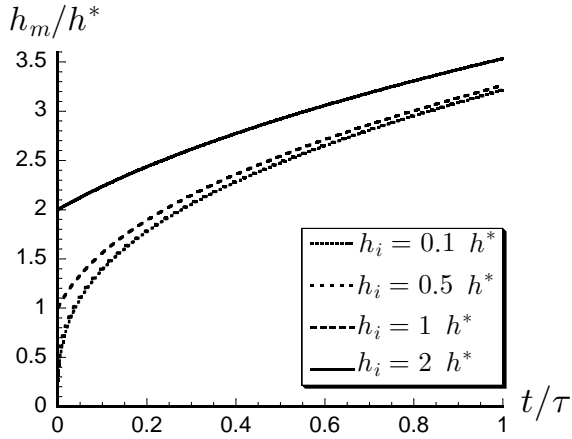} }
\caption{Short-time evolution of the rim height $h_{m}$: plot of $h_{m}$
versus $t/\tau$, for $h_{i}=$ 0.1, 0.5, 1 and 2 $h^{*}$. Note
that the curves corresponding to $h_{i}=$0.1 and 0.5$h^{*}$
cannot be distinguished within the graph scale.} \label{hm en t}
\end{figure}

The strain-rates $\alpha_m(t)$ and $\beta_m(t)$ at the rim edge
({\it i.e.} for $r=R_d(t)$) are simply related to the rim height
by balancing the radial stress acting on the rim and the capillary
forces. This gives the following equation:
\begin{equation}
\frac{h^*}{h_m} =
\frac{\alpha_m}{1+k\alpha_m^n}+\frac{\alpha_m-\beta_m}{1+k(\alpha_m-\beta_m)^n},
\label{alpham betam hm}
\end{equation}
admitting the approached forms $\beta_m \sim h^*/h_m$ for large
values of $h_m/h^*$, and $\beta_m \sim \mu(n,k) (h^*/h_m)^{1/1-n}$
for the thin films regime. $\mu(n,k)$ is a positive dimensionless
parameter depending on $n$ and $k$ as:
\begin{equation}
\mu(n,k) \equiv
\frac{k}{\psi(n)^{1-n}+(\psi(n)-1)^{1-n}}
\label{mu(n,k)}
\end{equation}
These expressions of strain-rates at the edge of the rim versus
the rim height enable us to find analytical expressions of the
growth laws during the initial stages of dewetting for thin films.

The rim height obeys the equation:
\begin{equation}
\frac{1}{h_m} \frac{\mathrm{D} h_m}{\mathrm{D} t} = \frac{\alpha_m -\beta_m} {\tau},
\label{hm}
\end{equation}
where $\mathrm{D} \equiv \frac{\dpt}{\dpt t}+v \frac{\dpt}{\dpt
r}$ stands for the particular derivative. In Fig.\ref{hm en t} we
present our results about the short-time evolution of rim height
for different initial thicknesses, with $t$ ranging from 0 to
$\tau$. The solution of Eq.\ref{hm} along with the initial
condition $h_m(t=0)=h_i$ is:
\begin{equation}
h_m(t) = h_i \left( 1+ \lambda(n,k,h_i) \frac{t}{\tau}
\right)^{1-n},
\label{hm cross}
\end{equation}
with $\lambda(n,k,h_i) \equiv k \frac{\psi^2-1}{\psi^{1-n}+1} (\frac{h^*}{h_i})^{1/1-n}$.
The adequation between our numerical results and this theoretical
prediction is excellent.

The dry radius is given by:
\begin{equation}
\frac{\dot{R_d}}{R_d}=\frac{\beta_m (t)} {\tau},
\label{rd}
\end{equation}
and we thus obtain:
\begin{equation}
R_d(t) = R_0 \left( 1+ \lambda(n,k,h_i) \frac{t}{\tau}
\right)^{\frac{1-n}{\psi-1}}
\label{rd cross}
\end{equation}
It is physically understandable that $R_{d}$, at a given time $t$,
is larger for thinner films : for a given applied force $|S|$ per
unit length, the thicker the film, the more the material to be
displaced and the lower the dewetting velocity.

Equations \ref{hm cross}-\ref{rd cross} allow one to define a
crossover time $t_c=\tau/\lambda(n,k,h_i)$. At short times ({\it
i.e.} $t \ll t_c$), $h_m$ and $R_d$ vary {\it linearly} with time:
$h_m \approx h_i (1+(1-n) t/t_c)$ and $R_d \approx R_0
(1+\frac{(1-n)}{\psi-1} t/t_c)$. At longer times ($t \gg t_c$),
the time dependence of $h_m$ and $R_d$ is a
{\it power-law}: $h_m \sim t^{1-n}$ and $R_d \sim
t^{(1-n)/(\psi-1)}$. Note that the thinner the film, the smaller
the crossover time $t_c$.

A simple expression of $R_d$ versus $h_m$ can be directly
derived from Eqs.\ref{hm cross} and \ref{rd cross}:
\begin{equation}
R_d(t) = R_0 \left( \frac{h_m(t)}{h_i}
\right)^{\frac{1}{\psi-1}}
\label{rd vs hm}
\end{equation}

\begin{figure}
\resizebox{0.45 \textwidth}{!}{%
\includegraphics*[1.2cm,9.5cm][19cm,23cm]{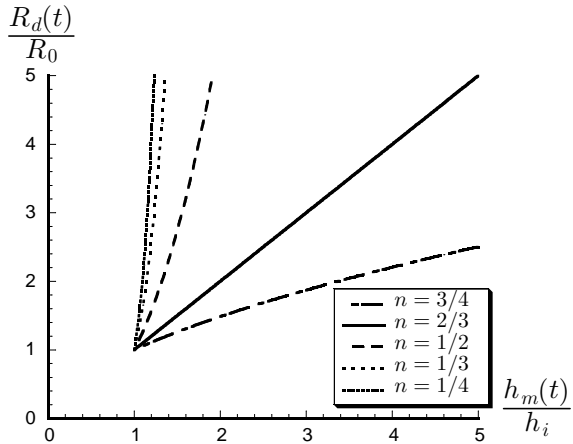} }
\caption{For thin films ($h_i \ll h^*$), during the initial stages
of growth, the dry radius is related to the rim height by a simple
power-law depending on the material rheology (Eq.\ref{rd vs hm}).
This figure shows the different regimes depending on $n$: for
$n=2/3$, the dependence is simply linear; for $n>2/3$, the radius
grows slower than the rim height, while for $n<2/3$, $R_d$ grows
fastly compared to the rim height increasing.}
\label{falrad}
\end{figure}

This relationship between the dry radius and the rim height is
shown in Fig.\ref{falrad} for different values of power-law index
$n$. Note that, for $n=2/3$, the expression (\ref{rd vs hm}) turns
out to be linear: this important property will be discussed in the
light of Reiter's experiments in part \ref{gunter exp}.

\subsubsection{Long-time regime of growth}
\label{thin final}

After the initial stages of hole growth, the driving force of the
phenomenon, $|S|$, is distributed over a rim height larger than
the characteristic thickness $h^*$. The resulting radial
constraint becomes weak, and gives rise to very small values of
strain-rates $\alpha$ and $\beta$. Then, to solve Eq.\ref{hm} in
this regime, we have to find the smallest-order correction to the
linear behavior of function $F$:
\begin{equation}
\alpha = F(\beta) \underset{\beta \ll 1}{\approx} \beta + \epsilon(\beta)
\label{def epsilon}
\end{equation}
Including this development in Eq.\ref{f(beta)}, and neglecting
terms involving second-order (or higher) $\epsilon$, we obtain a
non-linear first-order differential equation for $\epsilon$.
Solving it along with the condition $\epsilon(\beta=0)=0$ leads to
an explicit expression for $\epsilon(\beta)$. Keeping the
smallest order in development of $\epsilon$ for small $\beta$, we find \cite{non analyt}:
\begin{equation}
\alpha \underset{\beta \ll 1}{\approx} \beta + \left( \frac{n}{n+1}k \right)
\beta^{n+1}
\label{devt low beta}
\end{equation}

\begin{figure}
\resizebox{0.45 \textwidth}{!}{%
\includegraphics*[1cm,9.5cm][19.5cm,22cm]{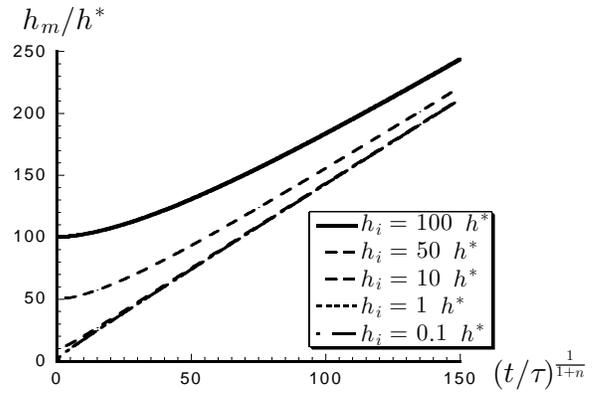} }
\caption{Long-time evolution of the rim height $h_{m}$, for
different initial thicknesses $h_i$ (0.1; 1; 10; 50; 100$h^*$),
with Cross law ($n=2/3$, $k=5$): plot versus
${(t/\tau)}^{\frac{1}{1+n}}$ to show the linear behavior of
$h_{m}$ with ${(t/\tau)}^{\frac{1}{1+n}}$ after the transient time
$t_{0}$. For a thick film ($h_{i} \gg h^{*}$), two regimes can be
distinguished. For $t < t_{0} \sim \tau
(\frac{h_{i}}{h^{*}})^{1+n}$, $h_{m}$ is constant and $R_{d}(t)$
increases exponentially with time. For $t> t_{0}$, $h_{m}$ growths
linearly with ${(t/\tau)}^{\frac{1}{1+n}}$. Note that the curves
corresponding to $h_{i}=$0.1 and 1$h^{*}$ cannot be distinguished
within the graph scale.} \label{hm en t^1+m}
\end{figure}

From Eq.\ref{hm}, it is then easy to obtain the following
expression for $h_m$:
\begin{equation}
h_m(t) = h_i \left( 1+ \nu(n,k,h_i) \frac{t}{\tau}
\right)^{\frac{1}{1+n}},
\label{hm cross long time}
\end{equation}
with $\nu(n,k,h_i) \equiv n k (\frac{h^*}{h_i})^{1+n}$. We have
checked numerically (cf. Fig.\ref{hm en t^1+m}) that this
power-law behavior is obeyed in the long-time regime of growth.

The corresponding law for dry radius in this regime is:
\begin{equation}
R_d(t) = R_0 \; \exp{\left[ \frac{1+n}{n k^2}\left(\frac{h_i}{h^*} \right)^n
\left((1+ \nu(n,k,h_i) \frac{t}{\tau})^{\frac{n}{1+n}}-1 \right) \right]}
\label{rd cross long time}
\end{equation}
For $t \gg \tau (h_i/h^*)^{1+n} /(nk)$, this law turns out to be a
stretched exponential:
\begin{equation}
R_d(t) \approx R_0 \; e^{\frac{1+n}{k} (nk)^{-\frac{1}{1+n}} \left(
\frac{t}{\tau} \right)^{\frac{n}{1+n}}}
\label{rd cross long time approx}
\end{equation}

From Eq.\ref{hm cross}, it is easy to define a crossover time
$t_e$ between the initial stages (part \ref{thin initial}) and the long-time regime of
growth (part \ref{thin final}):
\begin{equation}
t_e = \mu (\frac{h^*}{h_i})^{\frac{1}{1-n}} t_c
\label{crossover te}
\end{equation}

\subsubsection{Film profile}

\begin{figure}
\resizebox{0.5 \textwidth}{!}{%
  \includegraphics*[3.2cm,8.8cm][18.5cm,20.5cm]{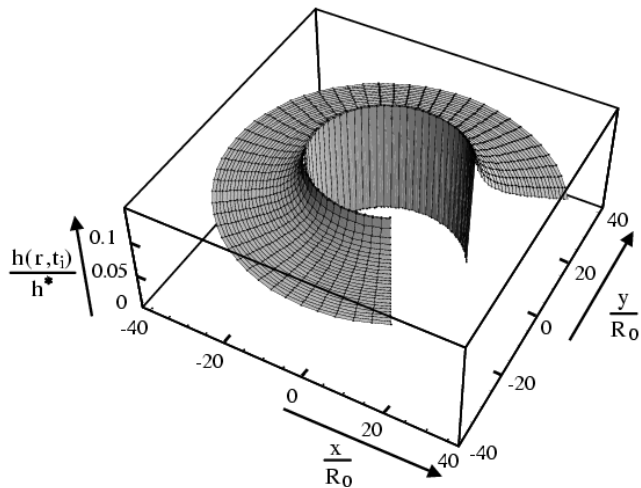} } \caption{Film profile
for $h_{i}=0.1 h^{*}$ and $t=2.10^{-2} \tau$. The constitutive
equation of the material is the Cross law (Eq.\ref{cross model}),
with a power-law index $n=2/3$. For clarity reasons, a sectional
view is displayed on a quarter of circle, and cartesian coordinates 
$(x,y,z)$,
naturally associated with cylindrical ones $(r,\Phi,z)$, are used.} \label{profil3d}
\end{figure}

\begin{figure}
\resizebox{0.47 \textwidth}{!}{%
  \includegraphics*[1.5cm,8.5cm][19.5cm,20.7cm]{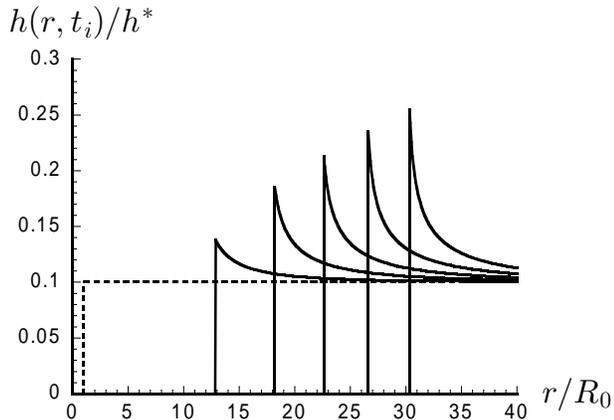} }
\caption{Evolution of a film (Cross model, $n=2/3$) profile
$h(r,t_i)$, for $h_{i}=0.1 h^{*}$ and $t_i$ ranging from $t=0$
(initial configuration, in dashed lines) to $t=5.10^{-2} \tau$
with a time step of $10^{-2} \tau$.} \label{profil qui monte}
\end{figure}

An example of profile obtained for $h_{i}=0.1 h^{*}$ is shown in
Fig.\ref{profil3d}. This profile is characterized by a highly
asymmetric shape for the rim. There is a striking similarity
between such a profile and those observed by Reiter for ultra-thin
films \cite{reiter}. On fig.\ref{profil qui monte} we present the
time evolution of this profile (for $t$ ranging from $t=0$ to
$t=5.10^{-2} \tau$, with a time step of $10^{-2} \tau$).

For the sake of conciseness, we focus there on the case
corresponding to the range of thicknesses and times covered by
Reiter's experiments, {\it i.e.} the initial times of growth for
thin films (cf. \cite{reiter} and \ref{discussion}). For clarity
reasons, let us write $\Omega(t) \equiv 1+ \lambda(n,k,h_i)
\frac{t}{\tau}$. In the regime $h_{i} \ll h^{*}$, the partial
derivative equation \ref{eqn hrt} for $h(r,t)$ reduces to:

\begin{equation}
\frac{\partial h(r,t)}{\partial t}+\frac{\kappa}{r^\psi}
\Omega(t)^{\frac{2-n-n\Psi}{\psi-1}}\frac{\partial h}{\partial r}
=(\psi-1)
\frac{\kappa}{r^{1+\psi}}\Omega(t)^{\frac{2-n-n\Psi}{\psi-1}} h
\label{edp profil}
\end{equation}

The resolution of Eq.\ref{edp profil} can be achieved using a
method of characteristics, as explained in appendix \ref{appa}.
The solution is:

\begin{equation}
h(r,t)= h_i
\frac{r^{\psi-1}}{(1+r^{\psi+1}-R_d(t)^{\psi+1})^{\frac{\psi-1}{\psi+1}}}
\label{solution profil}
\end{equation}

We checked (cf. appendix \ref{appa}) that this expression has a
wide range of validity: during all the initial growth, the
strain-rates remain high ($\alpha$ and $\beta \gg 1$) far away the hole
periphery, and thus Eq.\ref{solution profil} is an acceptable
description of the whole film profile.

\subsection{Thick films regime ($h_{i} \gg h^{*}$)}
For thick films, as shown in Fig.\ref{hm en t^1+m} (for instance in
the case $h_{i}=100 h^{*}$) two different time regimes can be
distinguished: at the beginning of hole formation, $h_{m}$ is
nearly constant and equals its initial value $h_{i}$ during a long
period of time, before growing faster. Here again, a distinction
between different time regimes is necessary.

\subsubsection{Initial stages of growth}
During the early stage, the strain rates at the
rim are constant and small, as a consequence of the constant large
thickness. In this case, the rheological law of the film is
viscous-type and the dry radius increases exponentially, as for
the case previously discussed (see Eq.\ref{reg visqueux}):

\begin{equation}
\left\{ \begin{array}{ll} R_{d}(t) = R_{0}
e^{\frac{|S|t}{\sigma_{0} \tau h_{i}} } = R_{0}
e^{\frac{t}{\tau_{i}}} \\
                          h_m(t) \approx h_i
                    \end{array}
            \right.
\end{equation}

This viscous-type behavior remains valid up to a crossover time
$t_{0} \approx \tau (\frac{h_{i}}{h^{*}})^{1+n}$.

\subsubsection{Long-time regime of growth}
We have checked numerically (cf. Fig.\ref{hm en t^1+m}) that the
anticipated analytical behavior
\begin{equation}
h_m(t) = h_i \left( 1+ \nu(n,k,h_i) \frac{t}{\tau}
\right)^{\frac{1}{1+n}},
\end{equation}
is obeyed after the crossover time $t_{0}$.

\begin{figure}
\resizebox{0.45 \textwidth}{!}{%
  \includegraphics*[1.2cm,8.5cm][20cm,21.5cm]{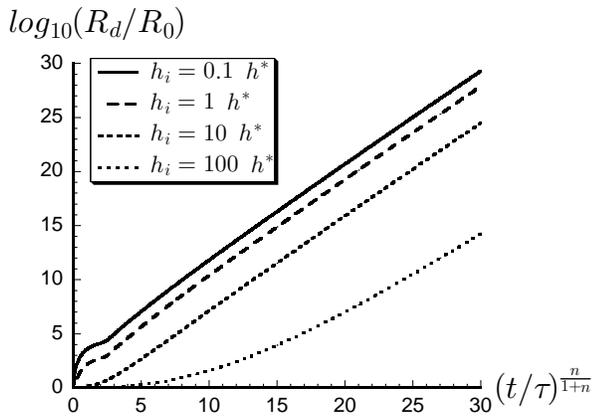} }
  \caption{Logarithmic plot
  of the dry zone radius $R_{d}$ versus $(t/\tau)^{n/(1+n)}$, for $h_{i}=$ 0.1, 1, 10 and 100
  $h^{*}$, in the case of Cross law with $n=2/3$ and $k=5$.
  Note the linear behavior of $\log_{10}{[R_{d}(t)/R_{0}]}$ with $(t/\tau)^{n/(1+n)}$
  after a transient time: the anticipated stretched exponential
  (Eq.\ref{rd cross long time approx}) is indeed obeyed.}
\label{rayon sec tpuiss}
\end{figure}

Similarly, the preceding exponential law for the dry radius
connects with the stretched exponential, whose power depends on
power-law index $n$ (cf. Eq.\ref{rd cross long time approx}):
\begin{equation*}
R_d(t) \approx R_0 \; e^{\frac{1+n}{k} (nk)^{-1/1+n} \left(
\frac{t}{\tau} \right)^{\frac{n}{1+n}}}
\end{equation*}
Fig.\ref{rayon sec tpuiss} which present the time evolution of
$R_{d}$ for different initial thicknesses $h_{i}$ with the Cross
model, shows that this anticipated behavior is obeyed for $t \gg
t_{0}$.

We emphasize the fact that the crossover from a simple exponential
to the stretched exponential regime is a consequence of the
non-linearity of rheological law \ref{rheolaw}.

\section{Discussion}
\label{discussion}

\subsection{Debregeas' experiments}

\begin{figure}
\resizebox{0.45 \textwidth}{!}{%
\includegraphics*[1cm,8.2cm][20cm,21cm]{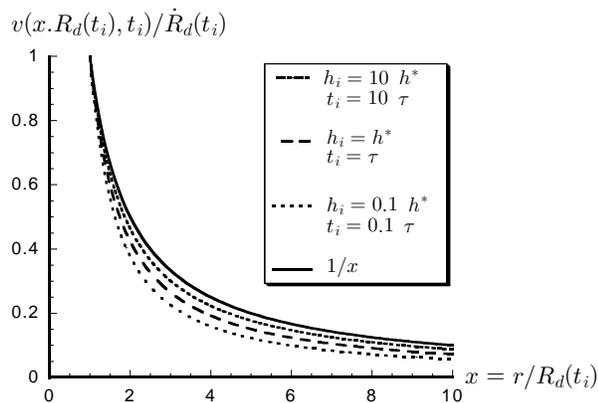} }
\caption{Velocity field $v(r,t_i)$ for different initial
thicknesses ($h_i=0.1 ; 1 ; 10 h^*$), at a given time $t_i$. All
radial distances are normalized by the dry radius at time $t_i$
for each thickness. A curve $1/x$ is represented for comparison.It
is clear that the velocity field is almost proportional to $1/r$
for a thick profile ($h_i=10 h^*$), similar to the long range
radial plug flow observed by Debr\'egeas.}
\label{speed}
\end{figure}

G. Debr\'egeas {\it et al.} \cite{debregeas} carried out
experiments on viscous bursting of freely suspended films of
long-chain polymers. The high molecular weights of
polydimethylsiloxane used in these experiments lead to high
viscosities ($\eta_0 \geq 600000$ cP) bursting processes where
viscous dissipation dominates inertia.

The range of thicknesses experimentally studied by Debr\'egeas
{\it et al.} (5 to 250 $\mu$m) covers the domain $h_{i} \gg
h^{*}$. As the corresponding crossover time is very large ($t_{0}
\gg \tau$), these experiments mainly covered the first regime $t
\ll t_{0}$ (the criterion $h_{i} \gg h^{*}$ is in fact a basic
hypothesis of their ``soft balloon'' model \cite{rem5}). In our
model, the velocity field $v(r,t)$ is found to be almost
proportional to $1/r$ at any time for thick profiles (cf.
Fig.\ref{speed}), similar to the experimentally observed long
range radial plug flow \cite{debregeas}.

Thus, our model accounts well for the exponential growth
and the absence of rim characteristics of the dewetting regime
observed by Debr\'egeas {\it et al.} for viscous polymer films.

\subsection{Reiter's experiments}
\label{gunter exp} G. Reiter \cite{reiter} studied the dewetting
of ultra-thin, almost glassy polystyrene films deposited onto
silicon wafers coated with a polydimethylsiloxane monolayer. The
thicknesses of the PS films used range from $h_i=$10 to $h_i=$60 nm.
Compared with the characteristic scale $h^* \approx 50$ nm, the
ratio $h_i/h^*$ ranges from about 0.2 to unity. The characteristic
time $\tau$ for PS under the experimental conditions is very long,
up to more than a year for the lowest temperatures conditions ($T$
ranges from $103^\circ$C to $130^\circ$C, to be compared with the
bulk glass transition temperature: $T_g=97.5^\circ$C). We can thus
expect that the relevant regime for comparing our results with
Reiter's experimental findings is the limit $h_i \ll h^*$, and
$t/\tau$ small (cf. \ref{thin initial}).

Three years ago, Dalnoki-Veress {\it et al.} made an experimental
study of hole formation and growth in freely standing PS films
\cite{dalnoki}. In agreement with Debregeas' experiments, they
observed exponential growth of the hole radius and uniform
thickening, but with a deviance from this regime for long-time
regimes of growth, a feature they analyzed in terms of
shear-thinning properties. They indeed characterized the polymer
rheology at the temperature of their experiments ($T= 115^\circ$C)
and found that the best fit for the viscosity $\eta$ versus the
shear strain rate $\dot{\gamma}$ is a power-law dependence:

\begin{equation}
\eta \sim |\dot{\gamma}|^{-d},
\end{equation}
with a power-law index $d=0.65 \pm 0.03$. For the following, let
us suppose that the rheology of ultra-thin PS films used by Reiter
can be described by the Cross model (Eq.\ref{cross model}) with
such a power-law index: $n=2/3$. It is of course an hypothesis
that would need an experimental validation. In particular, the
rheology of PS films at temperatures {\it very close} to the glass
transition may be quite different. Keeping this point in mind, the
following remarks are tentative explanations of some regimes of
dewetting experimentally observed by Reiter.

If $n=2/3$, Eq.\ref{def of psi} leads to the exact value $\psi=2$.
Then Eq.\ref{rd vs hm} turns out to be simply linear:

\begin{equation}
\frac{R_d(t)}{R_0}= \frac{h_m(t)}{h_i}
\label{linear}
\end{equation}

This property was indeed observed by Reiter: the maximum height of
the rim grows at the same speed as the radius of the hole
\cite{reiter}. The special role played by this value $n=2/3$ was
also pointed out by Shenoy and Sharma in their study
of the dewetting of power hardening viscoplastic solid \cite{shenoy}.

We can also focus on the individual dependences of $h_m$ and $R_d$
versus time. As precised in part \ref{thin initial}, for $t \gg
t_c$, the rim height and the hole radius follow a power-law time
dependence: $h_m \sim t^{1-n}$ and $R_d \sim t^{(1-n)/(\psi-1)}$.
In particular, for $n=2/3$, we obtain:
\begin{equation}
\left\{ \begin{array}{ll} h_m \sim (t/\tau)^{1/3} \\
                          R_d \sim (t/\tau)^{1/3}
                    \end{array}
            \right.
\label{power laws theor}
\end{equation}

Very recently, Reiter \cite{gunter} characterized the time
dependences of $R_d$ and $h_m$ for very thin films. He obtained
several dewetting regimes, and work is in progress to analyze
these results. It seems that several experimental factors must be
taken into account for this analysis. Among them, the influence of
film preparation appears to be decisive: the spin-cast PS film is
a highly metastable form of matter \cite{reiterpgg}. The sample
preparation (including the time of healing) can lead to thin films
where residual constraints on the chains persist, and may have an
influence on the dewetting process.

Quite interestingly, despite all these remarks, for a 20 nm PS
film deposited on a 10 nm PDMS layer around $T=120^\circ$C, the
results of Reiter's experiments seem to indicate that the maximum
height of rim and the hole radius obey power-laws:

\begin{equation}
\left\{ \begin{array}{ll} h_m \sim (t/\tau)^{a} & \mbox{with \;$a \approx 0.38 \pm 0.02$}\\
                          R_d \sim (t/\tau)^{b} & \mbox{with \;$b \approx 0.34 \pm 0.04$}
                    \end{array}
            \right.
\label{power laws exp}
\end{equation}

The time range on which Reiter calculated these power-law indexes
is up to about 100 minutes. As for a 20 nm film, the ratio
$h_i/h^*$ is smaller than unity, the crossover time $t_c$ is very
short, and we can expect that the power-law regime (Eq.\ref{power
laws theor}) is established. Thus we can compare Eqs.\ref{power
laws theor} and \ref{power laws exp} and remark that the two
experimental coefficients are indeed very close to the predicted
one: 1/3. This adequation between our model and Reiter's
experimental results appears to be satisfying. But many questions
remain open, as the role played by the thin PDMS monolayer on
which the PS film dewets is only taken into account by the free
slippage condition \cite{francoiseetpgg}. In fact, it was shown by
several studies that the lower layer is deformed by the moving
upper one \cite{wang}, which can leads to viscous dissipation and
to a marked decrease in the dewetting velocity \cite{lambooy}.

Let us conclude by a tentative idea: observing by TappingMode
atomic force microscopy (TM-AFM) the dry zone on the PDMS
substrate after the removal of PS film, Reiter
\cite{gunteraparaitre} discovered the formation of straight lines,
several microns long, starting to appear at hole diameters larger
than about 1 $\mu$m. These lines are PS microfibrils, whose width
can be as low as about 2 nm (to be compared with the radius of
gyration of the polymer in the bulk, of order 50 nm). This
observation suggests that the radial strain-rates are much larger
than the orthoradial ones at the hole periphery and involve this
elongational deformation of the molecules. This property could be
related to the fact that, in our model, the radial strain-rate
($\alpha$) is large compared with the orthoradial one ($\beta$)
during the initial stages of growth for thin films (cf.
Fig.\ref{function f(beta)}, regime $\beta \gg 1$). In particular,
note that this dissymmetry between values of $\alpha$ and $\beta$
is a direct consequence of the shear-thinning behavior of the
polymer: for a purely viscous fluid, $\alpha=\beta$ for any
thickness.

\section{Conclusion}

In conclusion, our model accounts well for the exponential growth
and the absence of rim characteristics of the dewetting regime
observed by Debr\'egeas {\it et al.} for viscous polymer films.
Taking into account the shear-thinning behavior of the polymers
near $T_{g}$ enables us to see the modifications induced by this
particular rheology on the film morphology. It appears from our
results that the early stages of dewetting for a shear-thinning
polymer film are mainly determined by the ratio of its initial
thickness $h_{i}$ to a characteristic scale $h^{*}$ (related to
the driving force of the process, $S$, and the rheological
response of the material, characterized by $\sigma_{0}$).

The film profile exhibits a sharp, asymmetric rim similar to the
one observed by Reiter in his latest experiments with ultrathin PS
films. The rich variety of growth behavior is summarized in
appendix \ref{bilan}. In all cases, the long-time evolution of
$h_m$ is a power-law (whose coefficient depends on the rheology of
the material studied), and $R_d$ obeys a stretched exponential
law. When taking for the power-law index $n$ the value obtained by
Dalnoki-Veress {\it et al.}, in the thin-film limit, our
predictions for the dry radius and the rim height time-evolutions
seem to explain some of Reiter's latest experiments. In
particular, the linear dependence of the rim height versus the dry
radius experimentally observed is predicted by our model for the
power-law index $n=2/3$.

Work is now in progress to incorporate in our model the
Laplace pressure (see \cite{rem4}) which might lead to oscillatory
dewetting fronts \cite{brenner,herminghaus}, and to evaluate the
possible effects of chain entanglements. In the future, we aim to
study the dewetting of thin polymer films {\it below} $T_{g}$,
where the existence of a yield stress in the rheological response
of the material may lead to new dewetting morphologies.

\section*{Acknowledgments}
We are grateful to {\sc G. Reiter} for very useful discussions and
for providing us with his experimental results on the dewetting of
ultra-thin polystyrene films prior to publication.

We also wish to acknowledge fruitful discussions with {\sc A. Aradian},
{\sc G. Debr\'egeas}, and {\sc M. Sferrazza}.

\appendix

\section{A few remarks about the velocity field inside the film}
\label{vz} Our choice of model (see \ref{the model}) for the
dewetting film deserves a few comments. As defined above, $v(r,t)$
is the radial part of flow field. The
incompressibility of the material implies the additional existence of a
small, but non-zero vertical component of the velocity field: let
us note $w$ this vertical part.

We already argued that the characteristic thickness of films
studied by Reiter and Debr\'egeas allows a {\it plug-flow}
assumption for the film: it implies that the radial part $v(r,t)$
is independent of variable $z$. This hypothesis does not imply the
same property for $w$: we must keep for it the most general
dependence $w(r,z,t)$.

In cylindrical coordinates, for a velocity field of the form $(v_r,
v_\Phi, v_z)$, the liquid incompressibility leads to:
\begin{equation}
\frac{\partial v_r}{\partial r}+ \frac{v_r}{r}+\frac{1}{r}
\frac{\partial v_\Phi}{\partial \Phi}+ \frac{\partial
v_z}{\partial z}=0 \label{general incomp}
\end{equation}

Here, as $v_\Phi=0$ for symmetry reasons, one has:
\begin{equation}
\frac{\partial v}{\partial r}+ \frac{v}{r}=-\frac{\partial
w}{\partial z} \label{incomp}
\end{equation}

Independence of $v$ versus $z$ allows a simple integration of
Eq.\ref{incomp} at a given $r$:

\begin{equation}
\int_{z=0}^{h(r,t)} (\frac{\partial v}{\partial r}+ \frac{v}{r})
\mbox{d}z=-\int_{z=0}^{h(r,t)}\frac{\partial w}{\partial z}
\mbox{d}z, \label{incomp integre}
\end{equation}
which gives:

\begin{equation}
h(r,t).(\frac{\partial v}{\partial r}+
\frac{v}{r})=-w\left(r,z=h(r,t),t\right) \label{integration faite}
\end{equation}

The vertical velocity at the top of the film, $w(r,z=h(r,t),t)$,
can be easily related to the particular derivative of $h(r,t)$,
provided that the film thickness remains small compared with the
characteristic lengths of variation of the radial speed component
and film thickness \cite{guyon hulin petit}:
\begin{equation}
w(r,z=h(r,t),t) \approx (\frac{\partial h}{\partial t}+
v.\frac{\partial h}{\partial r}) \label{approx hulin}
\end{equation}

Thus, in a more rigorous way, we recover the incompressibility
condition (see Eq.\ref{eqn hrt}):
\begin{equation*}
\frac{1}{h(r,t)} (\frac{\partial h(r,t)}{\partial t}+\frac{r
\beta(r,t)}{\tau} \frac{\partial h(r,t)}{\partial
r})=\frac{\alpha(r,t)-\beta(r,t)}{\tau}
\end{equation*}

The plug-flow assumption has a direct consequence: the left-hand
side of Eq.\ref{incomp} is independent of $z$; so does the
right-hand side, {\it i.e.} $\partial w /\partial z$, which
implies that $w$ has a simple linear dependence versus $z$:
\begin{equation}
w(r,z,t) = w(r,z=h(r,t),t) \frac{z}{h(r,t)} \label{linear}
\end{equation}

To finish with this point, note that taking into account
this {\it small} but {\it non-zero} vertical component of the
velocity field induces the presence of a supplementary term in the
conservation of momentum (equation \ref{motion}):

\begin{equation}
\frac{\partial \sigma_{rr}}{\partial r} +\frac{\partial
\sigma_{rz}}{\partial z} +\frac{\sigma_{rr}-\sigma_{\phi
\phi}}{r}=0 \label{motion bis}
\end{equation}

The possible consequences of this supplementary friction force on
our predictions for dewetting regimes will be studied in a
forthcoming publication.

\section{Results for logarithmic form of $\Phi$}
\label{log} For polymers, just above $T_{g}$, it is expected
within the framework of the free-volume model that $\sigma^{m}$
varies logarithmically with ${\dot{\gamma}}$ as \cite{dyre}:

\begin{equation}
\Phi(\dot{\gamma} \tau) = \ln(1+\dot{\gamma} \tau)
\label{shear-thinning}
\end{equation}

At low strain-rates ($\dot{\gamma} \tau <1$), this law displays a
viscous-type behavior ($\sigma^{m} \approx \eta_{0} \dot{\gamma}$,
with a zero-shear viscosity $\eta_{0}=\sigma_{0} \tau$), while for
large values of $\dot{\gamma} \tau$, $\sigma^{m}$ reaches an
almost constant value (shear-thinning behavior).

In order to deal with a minimum number of parameters, we assume
that $\sigma^{m} (\dot{\gamma})$ is an odd function
($\sigma^{m}(\dot{\gamma})= -\sigma_{0}\ln(1+|\dot{\gamma}| \tau)$
if $\dot{\gamma}<0$).

This appendix briefly presents the main results for thin and thick films.

{\it For thick films, at short times}, the dry radius and rim
height evolve as for a purely viscous fluid (cf. Eq.\ref{reg
visqueux}):

\begin{equation*}
\left\{ \begin{array}{ll} R_{d}(t) = R_{0}
e^{\frac{|S|t}{\sigma_{0} \tau h_{i}} } = R_{0}
e^{\frac{t}{\tau_{i}}} \\
                          h_m(t) \approx h_i
                    \end{array}
            \right.
\end{equation*}

{\it For thick films, at long times} ($t> t_0 \sim \tau
(h_i/h^*)^2$), the rim height grows proportionally to
$\sqrt{t/\tau}$:

\begin{equation}
h_{m}(t) \underset{t \gg t_{0}}{\sim} \frac{h^{*}}{2}
\sqrt{t/\tau},
\label{hm gd tps log}
\end{equation}

while the dry radius expands like
$\exp{(4\sqrt{t/\tau})}$:

\begin{equation}
R_{d}(t) \underset{t \gg t_{0}}{\sim} R_{\infty} e^{4
\sqrt{t/\tau}}.
\label{rayon sec a t g tau log}
\end{equation}

This law is also a stretched exponential, as for the Cross law.
This property is due to the fact that the function $\alpha =
F(\beta)$ is analytical (in the mathematical sense), {\it i.e.}
admits a series expansion near $0$ whose first non-linear term is
$\beta^2$. It indeed corresponds the limit behavior obtained by
taking the Cross law with $n \rightarrow 1$.

{\it For thin films, at short times}, a sharp rim grows very
quickly (with an initial speed $\dot{h}_{m} \approx h_{i}
\exp{h^{*}/2 h_{i}}$). In this regime, $h_{m}(t)$
is well described by the integral equation:

\begin{equation}
\int_{\frac{h^{*}}{2 h_{m}(t)}}^{\frac{h^{*}}{2 h_{i}}}
\frac{e^{-x}}{x} \mbox{d}x \underset{t \ll \tau}{\approx}
\frac{t}{\tau} \label{integral eqn}
\end{equation}

{\it For thin films, at long times}, as for the Cross law, the
hole growth connects with the long-time regime of thick films
(Eqs.\ref{hm gd tps log}-\ref{rayon sec a t g tau log}).

\begin{figure*}
\resizebox{0.7 \textwidth}{!}{%
\includegraphics*[0.7cm,14.2cm][20.2cm,25cm]{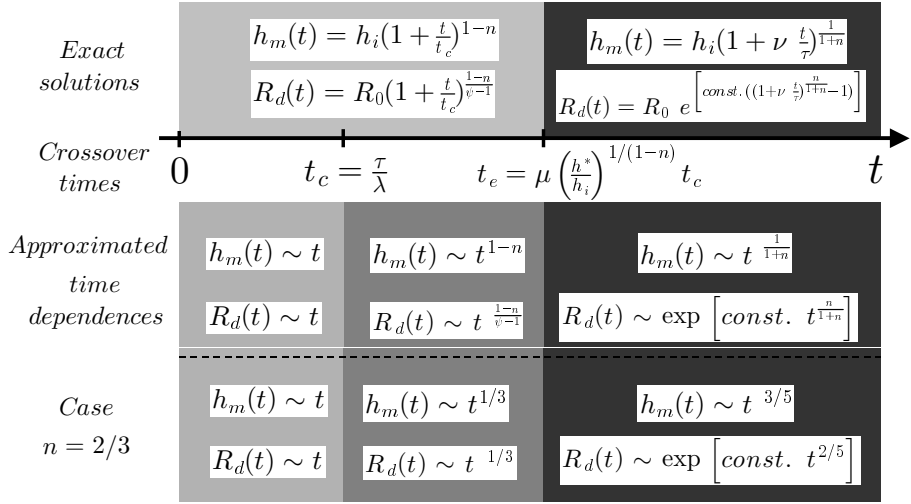} }
\caption{Summary of the different time regimes for the dewetting
of thin films ($h_i \ll h^*$). Depending on each time range, the
exact solutions and corresponding approximated time-dependences
are precised. We also point out the laws for the special value
$n=2/3$ (discussed in part \ref{gunter exp})} \label{hi ll ho}
\end{figure*}

\begin{figure*}
\resizebox{0.7 \textwidth}{!}{%
\includegraphics*[0.7cm,17cm][20.3cm,25cm]{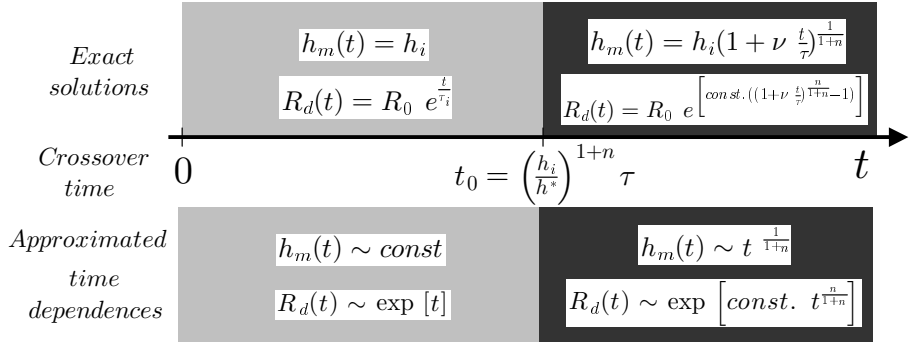} }
\caption{Summary of the different time regimes for the dewetting
of thick films ($h_i \gg h^*$). Depending on each time range, the
exact solutions and corresponding approximated time-dependences
are precised.} \label{hi gg ho}
\end{figure*}

\section{Summary of the dewetting regimes for a Cross constitutive
equation}
\label{bilan}

Figures \ref{hi ll ho} and \ref{hi gg ho} propose a summary of the
different dewetting time regimes for a Cross constitutive law, depending
on the initial thickness (to be compared with $h^*$), and relative
to different time ranges (to be compared with crossover times
$t_c$, $t_e$ and $t_0$).

\section{Analysis of the film profile}
\label{appa}

\subsection*{Method of characteristics}
To know the profile $h(r,t)$, the equation \ref{edp profil} has to
be solved:

\begin{equation*}
\frac{\partial h(r,t)}{\partial t}+\frac{\kappa}{r^{\psi}}
\Omega(t)^{\frac{2-n-n\Psi}{\psi-1}}\frac{\partial h}{\partial r}
=(\psi-1)
\frac{\kappa}{r^{1+\psi}}\Omega(t)^{\frac{2-n-n\Psi}{\psi-1}} h
\end{equation*}

along with the precedently precised initial condition
(Eq.\ref{boundary}) and the boundary values $h(r \rightarrow
\infty,t)=h_{i} \; \; (\forall t)$ and $h(r=R_d(t),t)=h_m(t)$. As
a a quasi-linear partial derivative equation, Eq.\ref{edp profil}
can be solved using a method of characteristics, as precised in
this appendix.

Writing $\dd h(r,t)=\frac{\partial h}{\partial t}\dd t +
\frac{\partial h(r,t)}{\partial r}\dd r$, we can replace the
partial derivative with respect to $t$ thanks to Eq.\ref{edp
profil}. Let define a characteristic by the function $r_{\xi}(t)$,
indexed by the parameter $\xi$ with $r_{\xi}(t=0)=\xi$, and
obeying the following ordinary differential equation:

\begin{equation}
\frac{\dd r_{\xi}(t)}{\dd t}= \frac{\kappa}{r_{\xi}(t)^{\psi}}
\Omega(t)^{\frac{2-n-n\psi}{\psi-1}} \label{def caract}
\end{equation}

Eq.\ref{def caract} is tractable, and gives each characteristic as
a function of time:
\begin{equation}
r_{\xi}(t)^{\psi+1}-\xi^{\psi+1}=
\Omega(t)^{\frac{(\psi+1)(1-n)}{\psi-1}}-1
\label{eqn caract}
\end{equation}

\begin{figure}
\resizebox{0.45 \textwidth}{!}{%
  \includegraphics*[3.5cm,9.2cm][20cm,21.5cm]{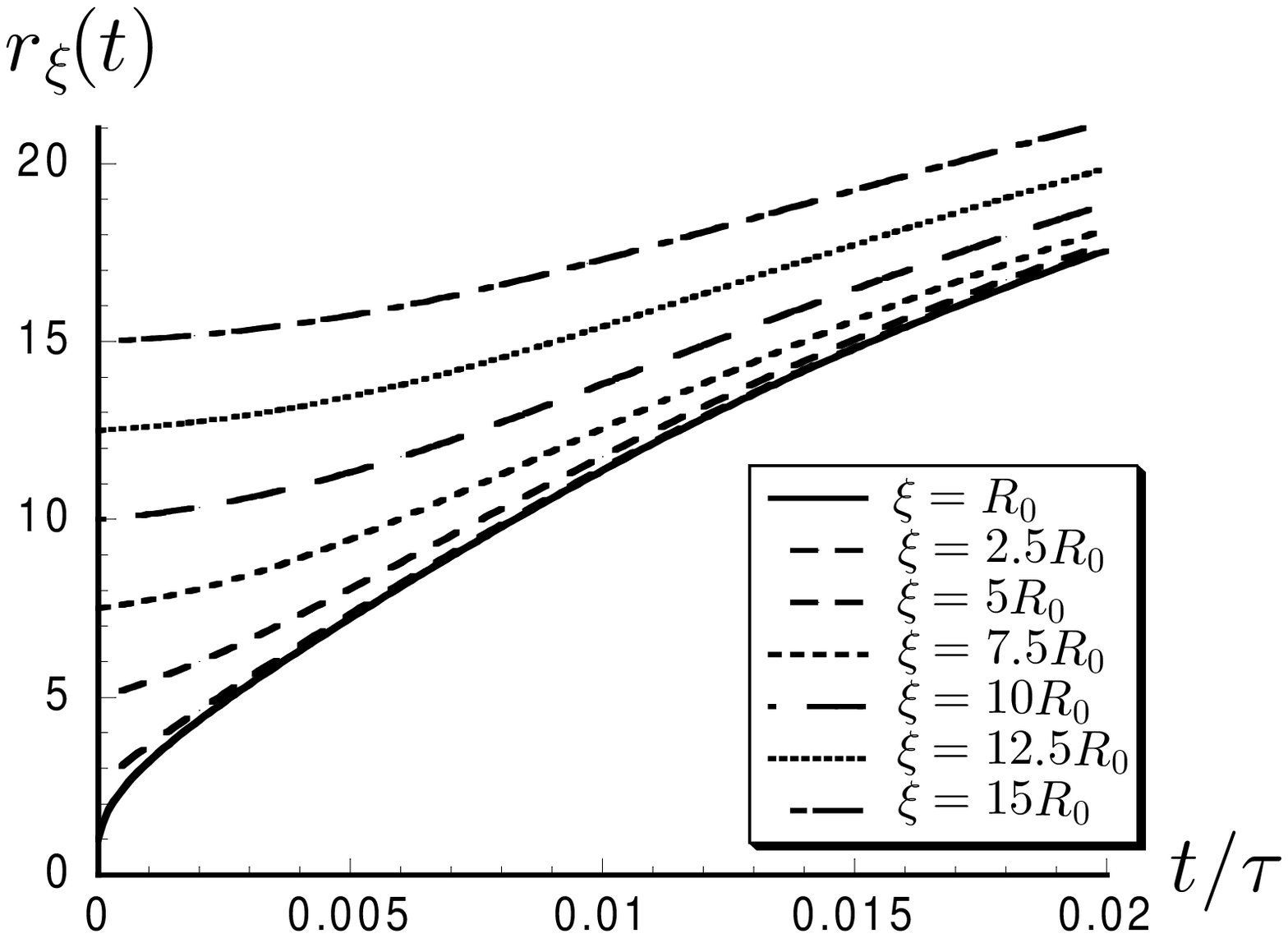} }
\caption{Graphic representation of characteristics $r_{\xi}(t)$,
for an initial thickness $h_i=0.1 h^*$ and $t$ ranging from 0 to
0.02$\tau$. Different values of $\xi$ parameter are taken:
$\xi=$1; 2.5; 5; 7.5; 10; 12.5 and 15$R_0$.}
\label{caracteristiques}
\end{figure}

In Fig.\ref{caracteristiques} is shown a representation of
characteristics $r_{\xi}(t)$ versus time, for a thin film
($h_i=0.1 h^*$). Note that these curves were extracted numerically
from our model, and are not a graphic illustration of solutions of
Eq.\ref{eqn caract}.

Along a given characteristic, the rim height $h_{\xi}(t)$ obeys a
simple equation:
\begin{equation}
\frac{\dd h_{\xi}(t)}{\dd t}= \frac{\kappa (\psi-1)}{r_{\xi}(t)^{\psi+1}}
\Omega(t)^{\frac{(\psi+1)(1-n)}{\psi-1}} h_{\xi}(t),
\label{eqn along char}
\end{equation}

whose solution is explicit and of the form:

\begin{equation}
h_{\xi}(t)= h_i \left(
\frac{\Omega(t)^{\frac{(\psi+1)(1-n)}{\psi-1}}+\xi^{1+\psi}-1}{\xi^{1+\psi}}
\right)^{\frac{\psi-1}{\psi+1}}
\label{hxit}
\end{equation}
Inverting Eq.\ref{eqn caract} to express $\xi$ as a function of
$r$ and $t$ enables us to extract the complete equation of the
profile (Eq.\ref{solution profil}):

\begin{equation*}
h(r,t)= h_i
\frac{r^{\psi-1}}{(1+r^{\psi+1}-R_d(t)^{\psi+1})^{\frac{\psi-1}{\psi+1}}}
\end{equation*}

\subsection*{Range of validity}
In the film zone near the hole ($r \gtrsim R_d(t)$), the
strain-rates $\alpha$ and $\beta$ are high. Thus Eq.\ref{eqn suppl}
gives:
\begin{equation}
\beta(r,t)=\beta_m(t) \left( \frac{R_d (t)}{r} \right)^{1+\psi} =
\kappa \frac{\Omega(t)^{\frac{2-n-n\psi}{\psi-1}}}{r^{1+\psi}}
\end{equation}

Our description of film profile remains valid as long as
$\beta(r,t) \geq 1$. Let us define the radial distance $r^*(t)$ up
to which this condition is verified : $\beta(r^*(t),t)=1$. The
ratio $r^*(t)/R_d(t)$ is given by:
\begin{equation}
\frac{r^*(t)}{R_d(t)}=
\kappa \Omega(t)^{\frac{1-n\psi}{\psi-1}}
\end{equation}

We regard the equation of profile as valid if $r^*(t) \gg R_d(t)$,
and arbitrarily choose a criterion $r^*(t) \geq \Upsilon R_d(t)$
with a factor $\Upsilon \sim 10$ for instance. It is then found
that $r^*(t) \geq \Upsilon R_d(t)$ as long as time $t$ is smaller
than a critical value $t^*$ given by:
\begin{equation}
t^*=\frac{1-n}{\psi-1} \frac{\mu(n,k)^{\frac{\psi (1-n)}{n\psi
-1}}}{\Upsilon^{\frac{\psi-1}{n \psi -1}}} \left( \frac{h^*}{h_i}
\right)^{\frac{\psi}{n\psi-1}}
\label{t*}
\end{equation}

For $h_i=0.1 h^*$, $n=2/3$, $k=5$, and $\Upsilon=10$, Eq.\ref{t*}
shows that the high strain-rates hypothesis remains valid on a
wide region as long as $t \lesssim 1000 \; \tau$. Thus, the
Eq.\ref{solution profil} is a good description of the film profile
during all the initial regime of growth.

\section{Influence of the initial shape of film profile}
\label{shape}
\begin{figure}
\resizebox{0.50 \textwidth}{!}{%
\includegraphics*[1.1cm,8.7cm][17.2cm,22cm]{smoot.eps} }
\caption{Evolution of the film morphology near the hole periphery
(from $R_d$ to $R_d+R_0$), starting with a continuous profile
(Eq.\ref{continuous} with $\chi=0.1$, filled with grey color).
Each profile, which is relative to the time $t_i$ and describes
the domain $r \geq R_d(t_i)$, is shifted to the point
$x=(r-R_d(t_i))/R_0=0$, so as to exhibit clearly its increasingly
asymmetric shape.} \label{smooth}
\end{figure}

The results presented above are obtained taking a discontinuous
shape for the initial profile of the film:
\begin{equation}
h(r,t=0)= \left\{ \begin{array}{ll} h_{i} & \mbox{if $r \geq R_{0}=1$} \\
                                    0 & \mbox{otherwise}
                    \end{array}
            \right.
\end{equation}

Does this discontinuity at $r=R_0$ play a role in the hole growth?
In other words, is it the source of the marked asymmetry observed
for the profiles obtained with our model?

To probe this hypothesis, we made the same resolution of our
system of equations, starting with a continuous profile:
\begin{equation}
h(r,t=0)= \left\{ \begin{array}{ll} h_{i}
e^{\frac{\chi}{(r/R_0)^2-1}} & \mbox{if $r > R_{0}=1$} \\
                                    0 & \mbox{if $r \leq R_{0}$}
                    \end{array}
            \right.
\label{continuous}
\end{equation}

This profile is infinitely derivable at $r=R_0$, the parameter
$\chi$ giving an order of the characteristic length of transition
between the regions $h=0$ and $h=h_i$. The profiles obtained with
this initial condition are shown in figure \ref{smooth}, with
$\chi=0.1$ (the shape of the initial profile, represented in
dashed lines in Fig.\ref{smooth}, is very smooth). The rim growth
seems to be very fast for a thin film (the time range of curves in
Fig.\ref{smooth} is 0 to $2.10^{-2}\tau$). Similarly, the
increasingly asymmetric shape of the profile shows that, even when
starting with a smooth profile, the resulting morphology of the
rim is the same as in Reiter's experiments. These considerations
prove that the discontinuity of the initial profile is {\it not} a
necessary condition for the appearance, growth and stiffening of
an asymmetric rim as those observed by Reiter.


\begin{thebibliography}{99}
\bibitem{rem1}
The driving force of the dewetting process is a difference of
interfacial tensions: $-S = \gamma_{SL}+\gamma_{LO}-\gamma_{SO}
>0$ (where SL corresponds to the solid/liquid interface, SO to
solid/air interface, etc...).
\bibitem{rem4}
In order to keep the model analytically tractable, we here neglect
the pressure discontinuity arising from the interface curvature.
Taking this Laplace pressure into account may lead to oscillations
of film profile, and may even give rise to a cascade of pinch-off
events, as proposed in Ref.\cite{condmat}.
\bibitem{non analyt}
The development of $F(\beta)$ at small $\beta$ shows that $F$
does not admit a Taylor expansion near $0$. This is directly due
to the fact that the Cross rheological equation \ref{cross model}
itself is not analytical, since the second term of its development
near $0$ is of order $n+1$.
\bibitem{rem5}
It is clear that this exponential law cannot be valid for too thin
films. Writing $\eta = E \tau_{rep}$ (where $E$ is an elastic
modulus, and $\tau_{rep}$ is the reptation time of polymers, the
relaxation time of the material), one obtains $\tau_{i}
=\tau_{rep} h/h_{0}$ with $h_{0} \equiv |S|/E$, which leads to a
characteristic time $\tau$ smaller than $\tau_{rep}$ if $h_{i} <
h_{0}$: the relaxation would be too fast.
\end{thebibliography}
\end{document}